\documentclass[11pt]{article}
\usepackage{amsthm}
\usepackage{amsfonts}
\usepackage{amsmath}
\usepackage{graphicx}
\usepackage{float} 

\usepackage{amssymb}
\usepackage{url}
\usepackage{color}

\begin{document}
	
	
	\newtheorem{theorem}{Theorem}
	\newtheorem{corollary}{Corollary}
	\newtheorem{proposition}{Proposition}
	\newtheorem{lemma}{Lemma}
	\newtheorem{definition}{Definition}
	\newtheorem{example}{Example}
	
	\newcommand{\TODO}[1]{\begingroup\color{red}#1\endgroup}
	\newcommand{\PFS}[1]{\begingroup\color{blue}#1\endgroup}

	\oddsidemargin 16.5mm
	\evensidemargin 16.5mm
	
	\thispagestyle{plain}

	
	
	
	
	\vspace{5cc}

	\begin{center}
		
		{\Large\bf
			Transit Functions and Clustering Systems
			\rule{0mm}{6mm}\renewcommand{\thefootnote}{}
			\footnotetext{\scriptsize ${}^{\ast}$Corresponding author. Manoj Changat}
			\footnotetext{
				
				\rule{2.4mm}{0mm}Keywords and Phrases.
				Transit function, Convexity, Binary clustering, Weak
				Hierarchies, Pyramids
		}}
		
		\vspace{1cc}
		{\it
			{Manoj Changat${}^{\ast}$, Ameera Vaheeda Shanavas, Peter F.\ Stadler}
		}
		
		\vspace{1cc}
		{\bf Abstract}\\
		\vspace{1cc}
		\parbox{24cc}{{\small
				Transit functions serve not only as abstractions of betweenness and
				convexity but are also closely connected with clustering systems.
				Here, we investigate the canonical transit functions of binary
				clustering systems inspired by pyramids, i.e., interval hypergraphs.
				We provide alternative characterizations of weak hierarchies, and
				describe union-closed binary clustering systems as a subclass of
				pyramids and weakly pyramidal clustering systems as an interesting
				generalization.
		}}
	\end{center}
	
	
	\vspace{1.5cc}
	\section{Introduction}
	
	Transit functions provide a unified framework to study intervals,
	convexities, and betweenness in graphs and posets
	\cite{Mulder:08}. Formally, a \emph{transit function} \cite{Mulder:08} on a
	finite non-empty set $X$ is a function $R: X\times X\to 2^X$ satisfying, for all
	$u,v\in X$, the three axioms
	\begin{description}\setlength{\itemsep}{0pt}\setlength{\parskip}{0pt}%
		\item[{(t1)}] $u\in R(u,v)$,
		\item[{(t2)}] $R(u,v)=R(v,u)$, and
		\item[{(t3)}] $R(u,u)=\{u\}$.
	\end{description}
	Transit functions were introduced originally to capture an abstract notion
	of betweenness in various discrete structures such as graphs, partially
	ordered sets, hypergraphs, etc. We say an element $x$ is between $u$ and
	$v$ if $x\in R(u,v)$.  On the other hand, a \emph{set system} on a
	non-empty finite set $X$ is any set $\mathcal{C}$ of subsets of $X$. In
	other words, set systems on $X$ can be considered as a hypergraph on $X$. A
	set system is called a \emph{clustering system} if $\emptyset \notin
	\mathcal{C}$, $X\in \mathcal{C}$ and every singleton set belongs to
	$\mathcal{C}$. Clustering or cluster analysis is a commonly used
	unsupervised approach in machine learning and data mining to aggregate
	data by similarity. Although often a partition of the base set or a
	hierarchy of clusters is desired, a much broad range of set
	systems that allow overlaps are of interest in community detection,
	see e.g.\ \cite{Girvan:02}. The systems of transit sets
	$\{R(u,v)|u,v\in X\}$ of transit functions $R$ also form clustering
	systems. This connection between transit function and clustering systems
	has been explored e.g.\ in
	\cite{Barthelemy:08,Bertrand:17,Changat:18a,Changat:19a}.
	
	This contribution is organized as follows. In section 2, we introduce
	basic properties of transit functions and set systems and review the
	relationship between binary clustering systems and their identifying
	transit functions. In section 3 we proceed to revisiting weak
	hierarchies and derive equivalent characterization for the clustering
	systems and their canonical transit functions. The following two
	sections use pyramidal set systems as a motivation for introducing
	properties of transit function that give rise to potentially interesting
	clustering systems. In section 4, we discuss a new covering property
	that yields an additional characterization of the transit functions
	of weak hierarchies. In section 5, we study union closed binary clustering
	systems and weakly pyramidal clustering systems as restriction and
	generalization of pyramids, respectively. Moreover, we further
	investigate some properties of transit functions that naturally appear in
	this context. Finally, we summarize the mutual implications of the
	properties studied here in Section 6.
	
	\vspace{1.5cc}
	\section{Background and preliminaries}
	
	Following \cite{Barthelemy:08}, a set system $\mathcal{C}\subseteq 2^X$ is
	\emph{pre-binary} if, for every $x,y\in X$, the collection of all clusters
	$C\in\mathcal{C}$ with $x,y\in C$ contains a unique inclusion-minimal
	member $C_{xy}$. These set systems are closely related to transit functions
	(called ``Boolean dissimilarities'' in \cite{Barthelemy:08}) that in
	addition satisfy the monotonicity axiom
	\begin{description}\setlength{\itemsep}{0pt}\setlength{\parskip}{0pt}%
		\item[(m)] $p,q\in R(u,v)$ implies $R(p,q)\subseteq R(u,v)$. 
	\end{description}
	The set systems corresponding to monotone transit functions were
	characterized in Proposition \ref{prop:1}\cite{Changat:19a} given in the
	following.
	
	A \emph{$\mathcal{T}$-system} is a system of non-empty sets
	$\mathcal{C}\subset 2^X$ satisfying the three axioms
	\begin{description}\setlength{\itemsep}{0pt}\setlength{\parskip}{0pt}%
		\item[(KS)] $\{x\}\in\mathcal{C}$ for all $x\in X$.
		\item[(KR)] For every $C\in\mathcal{C}$ there are points $p,q\in C$
		such that $p,q\in C'$ implies $C\subseteq C'$ for all
		$C'\in \mathcal{C}$.
		\item[(KC)] For any two $p,q\in X$ holds
		$\displaystyle \bigcap\{C\in\mathcal{C}|p,q\in C\} \in \mathcal{C}$.
	\end{description}
	
	\begin{proposition}[\cite{Changat:19a}]\label{prop:1}
		There is a bijection between monotone transit functions $R:X\times X\to
		2^X$ and $\mathcal{T}$-systems $\mathcal{C}\subseteq 2^X$ mediated by  
		\begin{equation}
			\begin{split}
				R_{\mathcal{C}}(x,y) &:= \bigcap\{ C\in\mathcal{C}| x,y\in C\}\\
				\mathcal{C}_R        &:= \{ R(x,y)| x,y\in X\}
			\end{split}
		\end{equation}
	\end{proposition}
	We say that a set system $\mathcal{C}$ \emph{is identified} by the transit
	function $R$ if $\mathcal{C}=\mathcal{C}_R$. A set system is called
	\emph{binary} \cite{Barthelemy:08}, if it is pre-binary and for every
	cluster $C\in\mathcal{C}$ there is a pair $p,q\in X$ such that
	$C=R_{\mathcal{C}}(p,q)$.  The latter condition and the definition of
	pre-binary are equivalent to (KR) and (KC), respectively.  In
	this case $R_{\mathcal{C}}$ is called the \emph{canonical transit function}
	of the set system $\mathcal{C}$, and $\mathcal{C}_R$ is the collection of
	\emph{transit sets} of $R$. Axiom (KS) corresponds to (t3), i.e.,
	the fact that all singletons are transit sets. The literature on clustering 
	typically assumes the base set $X$ itself is
	also a cluster. More precisely, a $\mathcal{T}$-system is a \emph{binary
		clustering system} if it satisfies
	\begin{description}\setlength{\itemsep}{0pt}\setlength{\parskip}{0pt}%
		\item[(K1)] $X \in \mathcal{C}$.
	\end{description}\vspace*{-\topsep}
	As shown in \cite{Barthelemy:08,Changat:19a}, binary clustering systems are
	identified by monotone transit functions satisfying the additional
	condition
	\begin{description}\setlength{\itemsep}{0pt}\setlength{\parskip}{0pt}%
		\item[(a')] There are $u,v\in X$ such that $R(u,v)=X$.
	\end{description}
	
	There is also a natural connection to convexities. A set system
	$(X,\mathcal{C})$ is \emph{closed (under non-empty intersection)} if it
	satisfies
	\begin{description}\setlength{\itemsep}{0pt}\setlength{\parskip}{0pt}%
		\item[(K2)] If $A,B\in\mathcal{C}$ and $A\cap B\ne\emptyset$ then
		$A\cap B\in\mathcal{C}$.
	\end{description}
	Note that convex set systems are conventionally defined to include the
	empty set $\emptyset$ \cite{Mulder:08}, in which case $(K2)$ could be
	simplified to ``$A,B\in\mathcal{C}$ implies $A\cap
	B\in\mathcal{C}$''. Clustering systems, however, by convention exclude
	$\emptyset$. That is, a set system $\mathcal{C}$ that satisfies (K1) and
	(K2) if and only if $\mathcal{C}\cup\{\emptyset\}$ is a convexity in the
	usual sense. For the purpose of this contribution, i.e., in the context of
	clustering systems, we will therefore refer to such a set system as a
	\emph{convexity}. Clearly, (K2) implies (KC).  Grounded convexities
	(satisfying (KS)) are therefore the same as \emph{closed clustering
		systems}.  It is shown in \cite{Changat:19a} that (K2) for $\mathcal{C}$ is
	equivalent to the following property of the identifying transit function:
	\begin{description}\setlength{\itemsep}{0pt}\setlength{\parskip}{0pt}%
		\item[(k)] For all $u,v,x,y\in X$ with $R(u,v)\cap R(x,y)\ne\emptyset$,
		there are $p,q\in X$ such that $R(u,v)\cap R(x,y) = R(p,q)$.
	\end{description}
	Axiom (k) was denoted (m') in \cite{Changat:18a,Changat:19a}.  Since the
	implication of (k) in particular implies $p,q\in R(u,v)\cap R(x,y)$, the
	axiom can be phrased equivalently as
	\begin{description}\setlength{\itemsep}{0pt}\setlength{\parskip}{0pt}%
		\item[(k)] For all $u,v,x,y\in X$ with
		$R(u,v)\cap R(x,y)\ne\emptyset$, there are $p,q\in R(u,v)\cap R(x,y)$
		such that $R(u,v)\cap R(x,y) = R(p,q)$.
	\end{description}
	\emph{Hierarchies} are classical clustering systems in which two clusters
	can not overlap. i.e., two clusters are either disjoint or one is properly
	contained in the other. Hierarchies are very restrictive as overlapping
	clusters occur in many real world natural clustering systems. Weak
	hierarchies, admitting overlapping clusters, were introduced as a
	generalization of hierarchies in \cite{Bandelt:89}.
	
	A clustering system $\mathcal{C}$ is a 
	\emph{weak hierarchy}
	\cite{Bandelt:89} if for any three sets $A,B,C\in\mathcal{C}$ holds
	\begin{equation}
		\label{eq:ABC}
		A \cap B \cap C \in \{ A\cap B, A\cap C, B\cap C\}
	\end{equation}
	The canonical transit
	functions of hierarchies were characterized axiomatically in
	\cite{Changat:18a} and that of weak hierarchies in \cite{Changat:19a}.  A
	convexity is a weak hierarchy if and only if its canonical transit function
	satisfies (w$_1$) \cite{Bertrand:17}:
	\begin{description}\setlength{\itemsep}{0pt}\setlength{\parskip}{0pt}%
		\item[(w$_1$)] There are no three distinct elements $x_1,x_2,x_3 \in X$
		such that for all $\{h,i,j\} = \{1,2,3\}$ holds $x_h\notin R(x_i,x_j)$.
	\end{description}
	Furthermore, $R$ is the transit function of a closed weak hierarchy if and
	only if it satisfies (m), (k), (a'), and (w$_1$) \cite{Changat:19a}.  The
	family of transit sets of the canonical transit function of a weak
	hierarchy is a convexity and hence closed \cite{Changat:19a}. A direct
	proof that (w) implies (k) for monotone transit functions can be found in
	\cite[Lemma 4.2]{Changat:19a}.  Moreover, (m) and (w$_1$) imply
	(a') \cite[Lemma.\ 6.4]{Changat:19a}, and in the presence of (m) and (a'),
	axioms (w$_1$) and the following axiom (w$_2$) are equivalent.
	\begin{description}\setlength{\itemsep}{0pt}\setlength{\parskip}{0pt}%
		\item[(w$_2$)] If $R(p,q)$, $R(u,v)$, and $R(s,t)$ have pairwise
		non-empty intersections, then $R(p,q)\cap R(u,v)\cap R(s,t) \in
		\{ R(p,q)\cap R(u,v),R(p,q)\cap R(s,t), R(u,v)\cap R(s,t) \}$.
	\end{description}
	
	Pyramids appear as clustering systems for data that admit a linear order:
	\begin{definition} \cite{Diday:86,Bertrand:13}
		A clustering system $(X,\mathcal{C})$ is \emph{pre-pyramidal} if there
		exists a total order $<$ on $X$ such that for every $C\in\mathcal{C}$ and
		all $x,y\in C$ we have $x<u<y$ implies $u\in C$. That is, all clusters
		$C\in\mathcal{C}$ are intervals w.r.t.\ $<$.
	\end{definition}
	We note in passing that in \cite{Bertrand:13} (K1) is replaced by the
	slightly weaker condition that $C\in\mathcal{C}$ is either a singleton or
	the union of proper subsets $C'\in\mathcal{C}$, $C'\subsetneq C$. A
	pre-pyramidal set system is \emph{pyramidal} if it is closed.  Naturally,
	we ask which transit functions $R$ correspond to pre-pyramidal clustering
	systems. A monotone transit function $R$ is (pre-)pyramidal if the family
	$\mathcal{C}_R$ of transit sets of $R$ forms a (pre-)pyramidal set
	system. Since pre-pyramidal set systems are weak hierarchies and the
	canonical transit functions of weak hierarchies satisfy (k), this is in
	particular also true for the canonical transit functions of pre-pyramidal
	clustering systems. Thus a monotone transit function $R$ is pre-pyramidal
	if and only if it is pyramidal.
	
	Pre-pyramidal set systems are also known as \emph{interval hypergraphs}
	\cite{Duchet:78}.  For any three distinct points $x,y,z\in X$, $y$ lies
	between $x$ and $z$ if every hyperpath connecting $x$ and $z$ has an edge
	containing $y$. A hypergraph $(X,\mathcal{C})$ is an interval hypergraph if
	and only if every set $\{x,y,z\}\subseteq X$ of three distinct vertices
	contains a vertex that lies between the other two
	\cite{Duchet:78}. Interval hypergraphs are characterized by an (infinite)
	series of forbidden induced sub-hypergraphs
	\cite{Tucker:72,Trotter:76,Duchet:84}.
	
	Following \cite{Eswaran:75,Nebesky:83} we define the union closure
	$\langle\mathcal{C}\rangle$ of an arbitrary set system
	$\mathcal{C}\subseteq 2^X$ as the set system containing all singletons and
	the unions of all pair-wisely intersecting $C',C''\in
	\mathcal{C}$. Nebesk{\'y} \cite{Nebesky:83} characterized interval
	hypergraphs, which he called \emph{projectoids}, in terms of the
	union-closure operation. The main result can be stated in our notation as
	follows: ``A set system $\mathcal{C}\subseteq 2^X$ is pre-pyramidal if and
	only if for any three $A,B,C\in\langle\mathcal{C}\rangle$ the union closure
	$\langle\{A,B,C\}\rangle$ is pre-pyramidal.'' The same paper also gives a
	convenient condition characterizing when $\langle\{A,B,C\}\rangle$ is
	pre-pyramidal.
	\begin{proposition}[\cite{Nebesky:83}]
		\label{prop:Nebesky2}
		Let $A,B,C\in 2^X$ be non-empty. Then $\langle\{A,B,C\}\rangle$ is
		pre-pyramidal if and only if the following two conditions are satisfied:
		\begin{description}\setlength{\itemsep}{0pt}\setlength{\parskip}{0pt}%
			\item[(W')] If $A\cap(B\setminus C)\ne\emptyset$ and
			$A\cap(C\setminus B)\ne \emptyset$ then $B\cap C\subseteq A$, and
			\item[(WP)] If $A$, $B$, $C$ have pair-wisely non-empty intersections,
			then one set is contained in the union of the two others.
		\end{description}
	\end{proposition}
	These known characterizations of interval hypergraphs do not seem to
	translate into simple, first order logic axioms for the corresponding
	transit functions. In the following, we, therefore, consider necessary and
	sufficient conditions.

	\vspace{1.5cc}
	\section{Weak hierarchies revisited}
	
	In this section, we briefly review on weak hierarchies and their
	identifying transit functions. Theorem \ref{thm:W} provides two additional
	characterizations of the canonical transit functions of weak hierarchies,
	and Lemma \ref{lem:w->x'} gives a necessary condition. Theorem
	\ref{lem:1=W} provides an additional characterization of weak
	hierarchies that will be useful later on.
	
	For a transit function on $X$, we consider the properties
	\begin{description}\setlength{\itemsep}{0pt}\setlength{\parskip}{0pt}%
		\item[(w)] For all $x,y,z\in X$ holds $z\in R(x,y)$ or $y\in R(x,z)$ or
		$x\in R(y,z)$.
		\item[(w$_2$)] $R(p,q)\cap R(u,v)\cap R(s,t) \in \{ R(p,q)\cap
		R(u,v),R(p,q)\cap R(s,t), R(u,v)\cap R(s,t) \}$.
		\item[(w$_3$)] If $x\in R(u,v)\setminus R(p,q)$, $y\in R(u,v)\cap R(p,q)$
		and $z\in R(p,q)\setminus R(u,v)$ then $y\in R(x,z)$.
	\end{description}  
	The characterizing axiom (w$_1$) of transit functions of
	weak hierarchies, is the contrapositive of (w). Axiom (w$_2$) is the
	translation of the most commonly used definition of a weak hierarchy into a
	transit axiom. Moreover, axiom (w$_3$) is obtained by observing the transit
	sets of monotone transit functions whose clustering systems $\mathcal{C}$
	are weak hierarchies. Clearly, linearly ordered intervals,
	i.e., interval hypergraphs, also share this property.
	\begin{theorem}
		\label{thm:W}
		Let $R$ be a monotone transit function on $X$.  Then (w), (w$_2$), and
		(w$_3$) are equivalent.
	\end{theorem}
	\begin{proof}
		(w$_2$) implies (w): Assume, w.l.o.g., $R(x,y)\cap R(y,z)\cap
		R(x,z)=R(x,y)\cap R(y,z)$. Since $y\in R(x,y)\cap R(y,z)$ we
		conclude $y\in R(x,z)$.
		\newline
		(w) implies (w$_3$): Suppose, for contradiction, that $R$ satisfies
		(w$_1$) but violates (w$_3$). Then there exist elements $u,v,x,y, p,q, z
		\in X$ with $R(u,v)\cap R(x,y)\ne\emptyset$ and $p\in R(u,v)\setminus
		R(x,y)$ and $q\in R(x,y)\setminus R(u,v)$, $z\in R(u,v)\cap R(x,y)$, but
		$z\notin R(p,q)$. Since $R$ is monotone by assumption, we have that
		$R(z,p)\subset R(u,v)$ and $R(z,q)\subset R(x,y)$. We also have $q\notin
		R(z,p)$ since otherwise it would imply that $q\in R(u,v)$, a
		contradiction to the assumption.  Similarly $p\notin R(z,q)$.  Therefore
		there exists distinct elements $p,q,z$ contradicting (w)
		\newline
		(w$_3$) implies (w): Suppose there are three distinct elements
		$x_1,x_2,x_3 \in X$ such that $x_1\notin R(x_2,x_3)$, $x_2\notin
		R(x_3,x_1)$ and $x_3\notin R(x_2,x_1)$. Starting from $x_2\in
		R(x_1,x_2)\cap R(x_2,x_3)$, (w$_3$) implies $R(x_1,x_2)\cap
		R(x_2,x_3)\subset R(p,q)$ for all $p\in R(x_1,x_2)\setminus R(x_2,x_3)$
		and $q\in R(x_2,x_3)\setminus R(x_1,x_2)$. Choosing $x_1$ as $p$ and
		$x_3$ as $q$ contradicts the assumption that $x_2\notin R(x_1,x_3)$.
		\newline
		(w$_3$) implies (w$_2$): To simplify the notation, set $C_1=R(p,q)$,
		$C_2=R(u,v)$, $C_3=R(s,t)$. If the intersection of any pair of transit
		sets in (w$_2$) is empty, then the intersection of all three of them is
		also empty. Suppose $C_1,C_2,C_3$ have pairwise non empty
		intersections. If one set, say $C_3$, is contained in another, say $C_2$,
		then $C_3\cap C_1\subseteq C_2\cap C_1$ and thus $C_1\cap C_2\cap
		C_3=C_1\cap C_3\ne\emptyset$. Now suppose none of the three sets is
		contained in another and consider the intersection of $C_3$ with $C_1\cup
		C_2$. We consider two cases: If $C_3\cap C_1 \subseteq C_3\cap C_2$ then
		$C_1\cap C_2\cap C_3=C_1\cap C_3\ne\emptyset$; similarly, $C_3\cap C_2
		\subseteq C_3\cap C_1$ implies $C_1\cap C_2\cap C_3=C_3\cap
		C_2\ne\emptyset$.  Otherwise, there are $x\in (C_3\cap C_1)\setminus
		C_2\subseteq C_1\setminus C_2$ and $z\in (C_3\cap C_2)\setminus
		C_1\subseteq C_2\setminus C_1$.  Now (w$_3$) implies $\emptyset\ne
		C_1\cap C_2\subseteq R(x,z) \subseteq C_3$, where the last inclusion
		follows by monotonicity, since $x,z\in C_3$. Thus $C_1\cap C_2\cap
		C_3=C_1\cap C_2$. Hence (w$_2$) holds.
	\end{proof}
	Lemma.\ 6.4 in \cite{Changat:19a} states that in the presence of (m) and
	(a'), axioms (w$_1$) and (w$_2$) are equivalent.  As the above theorem
	shows, the presence of (a') is not necessary for this result. Hence, each
	of these transit axioms (w$_2$) and (w$_3$) gives additional equivalent
	characterizations for transit functions of weak hierarchies.
	
	It is worth noting in this context that (w) does not imply
	monotonicity:
	\begin{example}
		Define transit function $R$ on $X= \{a,b,c,d,e\}$ as $R(a,b)= X$,
		$R(a,c)= \{a,c\}$, $R(a,d)= X$, $R(a,e)= \{a,e \}$, $R(b,c)=X$,
		$R(b,d)=\{b,e,d\}$, $R(b,e)= \{b,e\}$, $R(c,d)= \{a,c,d\}$, $R(c,e)=X$,
		$R(d,e)= \{d,e\}$. It is easy to verify that $R$ satisfies (w), but not
		(m) as $a,d\in R(c,d)$, and $R(a,d)=X\not \subset R(c,d)$.  It can be
		verified that $\mathcal{C}_R$ is not a weak hierarchy.
	\end{example}
	
	Before we proceed, we note that the following property, which appears in
	\cite{Changat:21a}\ in the context of cut-vertex transit functions, serves
	as generalization of (w).
	\begin{description}
		\item[(x')] If $z\notin R(x,y)$ then
		$R(x,y)\subseteq R(x,z)\cup R(z,y)$.
	\end{description}
	\begin{lemma}
		\label{lem:w->x'}
		A monotone transit function $R$ satisfying (w) also satisfies
		(x').
	\end{lemma}
	\begin{proof}
		By (w), $z\notin R(x,y)$ implies $x\in R(z,y)$ or $y\in R(z,x)$. In
		the first case we have $R(x,y)\subseteq R(z,y)$, and in the second case,
		we have $R(x,y)\subseteq R(z,x)$. In either case, therefore,
		$R(x,y)\subseteq R(x,z)\cup R(z,y)$. 
	\end{proof}
	The following example \ref{ex:x'notW} shows that converse need not be true.
	\begin{example}
		\label{ex:x'notW}
		Consider the monotone transit function $R$ on $X=\{a,b,c,d\}$ defined by
		$R(a,b) = \{a,b\}, R(a,c) = \{a,c\}, R(b,c) = \{b,c\}, R(a,d) = R(b,d)
		=R(c,d) = X$. $R$ satisfies (x') and violates (w).
	\end{example}
	
	In Proposition \ref{prop:Nebesky2}, the first condition, (W') turns out to 
	be yet another way of characterizing weak hierarchies:
	\begin{theorem}
		A set system $\mathcal{C}\subseteq 2^X$ satisfies Condition (W') in
		Proposition~\ref{prop:Nebesky2} if and only if $\mathcal{C}$ is a weak
		hierarchy.
		\label{lem:1=W}
	\end{theorem}
	\begin{proof}
		Consider three non-empty sets $A,B,C\in 2^X$ and assume that (W') is
		satisfied. If $B\cap C=\emptyset$ or $A\cap B=\emptyset$ or
		$A\cap C=\emptyset$, then the condition for weak hierarchy in
		equation.(\ref{eq:ABC}) is trivially satisfied. Now suppose $A$, $B$,
		$C$ intersect pairwisely. If $A\cap (B\setminus C)=\emptyset$ then
		$A\cap B\subseteq A\cap C$ and thus $A\cap B\cap C=A\cap B$. Similarly,
		if $A\cap (C\setminus B)=\emptyset$, then $A\cap B\cap C=A\cap C$.
		Otherwise (W') implies $B\cap C\subseteq A$ and thus
		$A\cap B\cap C=B\cap C$, and thus
		$A\cap B\cap C\in\{A\cap B, A\cap C, B\cap C\}$, i.e., $\mathcal{C}$ is a
		weak hierarchy.
		\newline
		To see the converse, assume that $\mathcal{C}$ is a weak hierarchy, $A$,
		$B$, and $C$ have pairwise nonempty intersection and assume
		$A\cap(B\setminus C)\ne\emptyset$ and $A\cap(C\setminus B)\ne
		\emptyset$. Thus there is $x\in A\cap B$, $x\notin C$ and thus
		$x\notin A\cap C$. Similarly, there is $y\in A\cap C$, $y\notin B$ and
		thus $y\notin A\cap B$. Therefore $A\cap B\cap C\ne A\cap B$ and
		$A\cap B\cap C\ne A\cap C$. Since $\mathcal{C}$ is a weak hierarchy,
		$A\cap B\cap C=B\cap C$ and thus $B\cap C\subseteq A$, i.e., Condition
		(W') of Proposition~\ref{prop:Nebesky2} is satisfied.  
	\end{proof}
	
	\vspace{1.5cc}
	\section{A covering property}
	
	In this section we introduce a rather general covering property (mm)
	of transit functions inspired by an order on $X$.  Using (mm) and another
	axiom (x'), which is already weaker than (w), we derive, in Theorem
	\ref{thm:w=mm+x'}, an alternative characterization of the canonical transit
	functions of weak hierarchies. Moreover, we briefly discuss certain
	generalizations.
	
	Consider two overlapping intervals $[x,y]$ and $[u,v]$ with $x<u\le y<v$
	and their union $[x,v]$ w.r.t.\ some order $<$ on $X$. A monotone transit
	function compatible with this order must satisfy $R(x,y)\cup
	R(u,v)\subseteq R(x,v)$. If the order is unknown, however, we only know
	that, unless $R(u,v)\subseteq R(x,y)$ or $R(x,y)\subseteq R(u,v)$, there
	must be points $p\in R(u,v)\setminus R(x,y)$ and $q\in R(x,y)\setminus
	R(u,v)$ such that the transit set $R(p,q)$ contains both $R(x,y)$ and
	$R(u,v)$. The fact that in this case $p$ and $q$ cannot both be located in
	either $R(u,v)$ and $R(x,y)$, is a simple consequence of monotonicity.
	This simple reasoning suggests to consider the following property:
	\begin{description}\setlength{\itemsep}{0pt}\setlength{\parskip}{0pt}%
		\item[(mm)] If $R(x,y)\cap R(u,v)\ne\emptyset$ then there are
		$p,q\in R(x,y)\cup R(u,v)$ such that
		$R(x,y)\cup R(u,v) \subseteq R(p,q)$.
	\end{description}
	Note that for monotone transit functions the implication of $(mm)$ is
	trivially true with $\{p,q\}=\{u,v\}$ whenever $R(x,y)\subseteq
	R(u,v)$. Property (mm) has a simple translation to the corresponding
	$\mathcal{T}$-system:
	\begin{description}\setlength{\itemsep}{0pt}\setlength{\parskip}{0pt}%
		\item[(MM)] For every $C',C''\in \mathcal{C}$ with
		$C'\cap C'' \neq \emptyset$ there are $p,q\in C'\cup C''$ such that
		the inclusion-minimal set $C_{pq}$ containing $p$ and $q$ satisfies
		$C'\cup C''\subseteq C_{pq}$.
	\end{description}
	
	\begin{lemma}
		A transit function $R$ satisfying (mm) also satisfies (a').
	\end{lemma}
	\begin{proof}
		Suppose $R(u,v)\ne X$, i.e., there is $w\notin R(u,v)$. Since
		$R(u,v)\cap R(v,w)\ne\emptyset$, (mm) implies
		$R(u,v)\cup R(v,w)\subseteq R(u',v')$ for some $u',v'$, and thus
		$|R(u',v')|>|R(u,v)|$. Since $X$ is finite, repeating this construction
		will eventually reach two points $u_k,v_k$ such that $R(u_k,v_k)=X$. 
	\end{proof}
	
	\begin{figure}[t]
		\begin{minipage}{0.20\textwidth}
			\includegraphics[width=\textwidth]{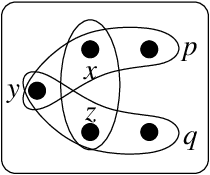}
		\end{minipage}  \quad\qquad
		\begin{minipage}{0.70\textwidth}
			\caption{The set system $\mathcal{C}$ on $X$ comprising the
				singletons, $X$, the edge $A=\{x,z\}$, and the two triples
				$B=\{x,y,p\}$ and $C=\{y,z,q\}$ is not a weak hierarchy
				since $A$, $B$, and $C$ intersect pair-wisely but $A\cap B\cap
				C=\emptyset$. On the other hand, it is a binary clustering system
				and satisfies (MM).}
			\label{fig:exMM}
		\end{minipage}
	\end{figure}
	
	Hence a $\mathcal{T}$-system satisfying (MM) is a binary clustering
	system. Despite the motivation for introducing (mm), this property captures
	very little of ordering structure. In fact, (mm) turns out to be a proper
	generalization of (w):
	\begin{lemma}
		\label{lem:w->mm}
		Let $R$ be a monotone transit function. Then (w) implies (mm).
	\end{lemma}
	\begin{proof}
		If $R(x,y)\subseteq R(u,v)$ or $R(u,v)\subseteq R(x,y)$, the claim holds
		trivially for $p,q=u,v$ or $p,q=x,y$, respectively. Thus, suppose there
		are $p_1\in R(x,y)\setminus R(u,v)$ and $q_1\in R(u,v)\setminus R(x,y)$.
		By (w$_3$), $R(p_1,q_1)$ contains $R(u,v)\cap R(x,y)$. If
		$R(u,v)\subseteq R(p_1,q_1)$ and $R(x,y)\subseteq R(p_1,q_1)$, we are
		done. Otherwise, there is $p_1'\in R(x,y)\setminus R(p_1,q_1)$ or
		$q_1'\in R(u,v)\setminus R(p_1,q_1)$. We proceed by showing that $p_1'$
		and $q_1'$ can be used to expand $R(p_1,q_1)$:
		\newline
		Assume that $p_1'$ exists. Then (w) implies $p_1\in R(q_1,p_1')$ or
		$q_1\in R(p_1,p_1')$. Since $p_1,p_1'\in R(x,y)$ and $q_1\in
		R(u,v)\setminus R(x,y)$, the second alternative is ruled out. Now (m)
		implies $R(p_1,q_1)\subset R(p_1',q_1)$. The inclusion is strict since,
		by construction, $p_1'\notin R(p_1,q_1)$. If $p_1'$ does not exist, an
		analogous argument shows that $R(p_1,q_1)\subset R(p_1,q_1')$.  If $p_1'$
		exists, we set $(p_2,q_2):=(p_1',q_1)$, otherwise we use
		$(p_2,q_2):=(p_1,q_1')$.
		\newline
		The construction of the previous paragraph can be repeated to generate a
		sequence of pairs $(p_i,q_i)$ such that $p_i=p_{i-1}$ and $q_i\in
		R(u,v)\setminus R(p_{i-1},q_{i-1})$ or $q_i=q_{i-1}$ and $p_i\in
		R(x,y)\setminus R(p_{i-1},q_{i-1})$ for which $R(p_{i-1},q_{i-1})\subset
		R(p_i,q_i)$. Since the cardinality $|R(p_i,q_i)\cap (R(x,y)\cup R(u,v))|$
		strictly increases with $i$ due to the addition of $p_i$ or $q_i$, the
		sequence must terminate at a finite $k$, namely the smallest integer for
		which neither $p_k'$ nor a $q_k'$ exists. By construction we then have
		$R(x,y)\cup R(u,v) \subseteq R(p_k,q_k)$.
	\end{proof}
	
	Example~\ref{ex:MMnotW} shows that there are clustering systems satisfying
	(MM) that are not weak hierarchies:
	\begin{example}
		\label{ex:MMnotW}
		The clustering system $(X,\mathcal{C})$ in Figure.~\ref{fig:exMM} has the
		canonical transit function $R(x,y)=R(x,p)=R(y,p)=\{x,y,p\}$,
		$R(y,z)=R(y,q) = R(z,q)=\{y,z,q\}$, $R(x,z)=\{x,z\}$ and
		$R(p,z) = R(p,q)= R(x,q)= X$. One easily checks that $R$ is monotone and
		satisfies (mm). However, $R(x,z)$, $R(x,y)$, and $R(y,z)$ violate  
		(w).
	\end{example}
	Examples~\ref{ex:x'notW} and~\ref{ex:MMnotW} show that properties (mm)
	and (x') are independent for monotone transit function.
	\begin{theorem}
		\label{thm:w=mm+x'}
		A monotone transit function $R$ satisfies (w) if and only if it
		satisfies (mm) and (x').
	\end{theorem}
	\begin{proof}
		If $R$ satisfies (w), then (x') and (mm) holds by Lemmas~\ref{lem:w->x'}
		and~\ref{lem:w->mm}, respectively. For the converse suppose (x') and (mm)
		are satisfied but (w) does not hold.  Then there are three points
		$a,b,c\in X$ such that $a\notin R(b,c), b\notin R(a,c)$ and $c\notin
		R(a,b)$. Since $R(a,b)\cap R(a,c)\neq\emptyset$, (mm) implies that there
		are $p,q\in R(a,b)\cup R(a,c)$ such that $R(a,b)\cup R(a,c)\subseteq
		R(p,q)$.  If $(p,q)=(b,c)$ then $R(a,b)\cup R(a,c)\subseteq R(b,c)$,
		which implies $a\in R(b,c)$, a contradiction.  If $(p,q)\neq(b,c)$ then
		at least two of the sets $R(a,b)$, $R(b,c)$, and $R(a,c)$ contain more
		than two points, and hence there exists $d\in R(a,b)\cup R(a,c)$ such
		that $d\neq a,b,c$.
		\newline
		Now suppose $p=d$ and $q\in\{a,b,c\}$. W.l.o.g.\ we may assume that $d\in
		R(a,b)$. If $(p,q)=(d,a)$, (m) implies $R(p,q)=R(d,a)\subseteq R(a,b)$, a
		contradiction. Similarly, $(p,q)=(d,b)$ implies $R(p,q)=R(d,b)\subseteq
		R(a,b)$, again a contradiction.  \newline
		It remains to consider $(p,q)=(d,c)$. Here we have, (*) $R(a,b)\cup
		R(a,c)\subseteq R(d,c)$.  Since $c\notin R(a,b)$, (x') implies
		$R(a,b)\subseteq R(a,c)\cup R(b,c)$ and thus $d\in R(a,b)$ yields $d\in
		R(a,c)$ or $d\in R(b,c)$.  If $d\in R(a,c)$ then $R(d,c)\subseteq
		R(a,c)$, and (*) implies $b\in R(a,c)$, a contradiction. If $d\in R(b,c)$
		then $R(d,c)\subseteq R(b,c)$ and (*) implies $a\in R(b,c)$, a
		contradiction.  Thus $q\notin\{a,b,c\}$ and there must be a point $ e \in
		R(a,b)\cup R(a,c)$ such that $e\neq a,b,c,d$.  \newline
		If $(p,q) = (d,e)$ then (**) $R(a,b)\cup R(a,c)\subseteq R(d,e)$.  If
		$d,e\in R(a,b)$ or $d,e\in R(a,c)$, then (m) contradicts (**).  W.l.o.g.,
		we may assume $d\in R(a,b)\setminus R(a,c)$ and $e\in R(a,c)\setminus
		R(a,b)$. Since $d\notin R(a,c)$, (x') implies $R(a,c)\subseteq R(a,d)\cup
		R(d,c)$ and $e\in R(a,c)$ implies $e\in R(a,d)$ or $e\in R(d,c)$. If
		$e\in R(a,d)$, then $R(d,e)\subseteq R(a,d)\subseteq R(a,b)$ and (**)
		implies $c\in R(a,b)$, a contradiction.  If $e\in R(d,c)$, then
		$R(d,e)\subseteq R(d,c)$ and $R(a,b)\cup R(a,c)\subseteq R(d,c)$, a
		situation that we have already ruled out in the previous paragraph. Thus
		no three points $a,b,c$ violating (w) can exist.
	\end{proof}
	
	Property (MM) suggests a further relaxation to the simple covering
	property:
	\begin{description}\setlength{\itemsep}{0pt}\setlength{\parskip}{0pt}%
		\item[(K3)] For every $C',C''\in\mathcal{C}$ with $C'\cap C''\neq
		\emptyset$ there is a unique inclusion-minimal set $C$ that satisfies
		$C'\cup C''\subseteq C$.
	\end{description}
	It can be translated to the language of transit functions as follows:
	\begin{description}\setlength{\itemsep}{0pt}\setlength{\parskip}{0pt}%
		\item[(k3)] If $R(x,y)\cap R(u,v)\ne\emptyset$ then there are $p,q\in X$
		such that (i) $R(x,y)\cup R(u,v)\subseteq R(p,q)$ and (ii) if
		$R(x,y)\cup R(u,v)\subseteq R(p',q')$ then $p,q\in R(p',q')$.
	\end{description}
	We note that relaxing property (k3) further, requiring only the first
	property is equivalent to (a'), since the fact that $X$ appears as a
	transit set is sufficient to satisfy condition (i). The following
	examples \ref{ex:MMK3top} and \ref{ex:MMK3bottom} show that neither of 
	the implications (MM) $\implies$ (K3) $\implies$ (K1) can be reversed.
	
	\begin{figure}[t]
		\begin{minipage}{0.20\textwidth}
			\includegraphics[width=\textwidth]{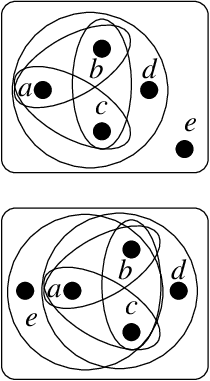}
		\end{minipage} \quad\qquad
		\begin{minipage}{0.70\textwidth}
			\caption{Top: the set system $\mathcal{C}$ on $X$ comprising the
				singletons, $X$, the three edges $\{a,b\}$, $\{a,c\}$, $\{b,c\}$,
				and the set $\{a,b,c,d\}$ satisfies (KS), (KR),
				(KC). It satisfies (K3) since the three pairs of
				overlapping edges are contained in the unique inclusion-minimal
				cover $\{a,b,c,d\}$. It does not satisfy (MM) since the
				inclusion minimal set containing $p,q\in\{a,b\}\cup\{b,c\}$ is
				always one of the three edges. \newline Bottom: the set system
				$\mathcal{C}$ on $X$ comprising the singletons, $X$, the three
				edges $\{a,b\}$, $\{a,c\}$, $\{b,c\}$, and the two sets
				$B_1=\{a,b,c,d\}$ and $B_2=\{a,b,c,e\}$ satisfies (KS),
				(KR), (KC), and (K1). It violates (K3) since
				both $B_1$ and $B_2$ are inclusion-minimal covering sets for
				$\{a,b\}\cup\{b,c\}$.}
			\label{fig:exMMK3}
		\end{minipage}
	\end{figure}
	
	\begin{example}
		\label{ex:MMK3top}
		The set system $\mathcal{C}$ shown in Figure.~\ref{fig:exMMK3}(top) is a
		$\mathcal{T}$-system and satisfies (K3) since the three pairs of
		overlapping edges are contained in the unique inclusion-minimal cover
		$\{a,b,c,d\}$. It violates (MM) since the inclusion minimal set
		containing $p,q\in\{a,b\}\cup\{b,c\}$ is one of the three edges.
	\end{example}
	
	\begin{example}
		\label{ex:MMK3bottom}
		The set system $\mathcal{C}$ shown in Figure.~\ref{fig:exMMK3}(bottom)
		comprising the singletons, $X$, the three edges $\{a,b\}$, $\{a,c\}$,
		$\{b,c\}$, and the two sets $B_1=\{a,b,c,d\}$ and $B_2=\{a,b,c,e\}$ is a
		$\mathcal{T}$-system and satisfies (K1) since $X\in\mathcal{C}$.  It
		violates (K3) since both $B_1$ and $B_2$ are inclusion-minimal
		covering sets for $\{a,b\}\cup\{b,c\}$.
	\end{example}
	
	\begin{lemma}
		Let $(X,\mathcal{C})$ be a $\mathcal{T}$-system satisfying (K2).
		Then (K1) implies (K3).
	\end{lemma}
	\begin{proof}
		Consider two sets $C',C''\in\mathcal{C}$.  Since $X\in\mathcal{C}$, there
		is a covering set $C$ with $C'\cup C''\subseteq C$. Now denote by
		$C_1,C_2,\dots,C_k$ all sets $C_i\in\mathcal{C}$ with $C'\cup
		C''\subseteq C_i$. By (K2) the intersection $C^*=\bigcap_k C_k
		\in\mathcal{C}$, and thus $\mathcal{C}$ harbours a unique covering set
		for $C'\cup C''$.
	\end{proof}
	Example~\ref{ex:MMK3top} also satisfies (K2), hence (K3) does not
	imply (MM) even in a convexity.

	\vspace{1.5cc}
	\section{Union-closure and weakly pyramidal clustering} 
	
	The union of two overlapping intervals is again an interval. This simple
	observation suggests to consider the \emph{union-closure} property:
	\begin{description}\setlength{\itemsep}{0pt}\setlength{\parskip}{0pt}%
		\item[(uc)] If $R(x,y)\cap R(u,v)\ne\emptyset$ then there are
		$p,q\in R(x,y)\cup R(u,v)$ such that
		$R(x,y)\cup R(u,v)=R(p,q)$.
	\end{description}
	As we shall see below, it implies several interesting weaker
	properties. It explains union-closed binary clustering systems, which
	turn out to be a proper subclass of pyramids. We then consider natural
	relaxations, which leads us to the notion of weakly
	pyramidal transit functions, a proper subclass of weak hierarchies that
	properly contains the pyramids.
	
	We start by noting that (uc) does not hold for all pyramidal set
	system:
	\begin{example}
		\label{ex:pynotuc}
		Consider the binary clustering system $(X,\mathcal{C})$ with
		$X=\{1,2,\dots,n\}$ with $n\ge 4$ such that $\mathcal{C}$ comprises the
		singletons, $X$, and all edges $\{k,k+1\}$ for $1\le k<n$. Clearly
		$(X,\mathcal{C})$ is pyramidal and but not union-closed, since
		$\{1,2\}\cup\{2,3\}=\{1,2,3\}\notin\mathcal{C}$.
	\end{example}
	
	\begin{lemma}
		\label{lem:uc->k}
		A monotone transit function $R$ satisfying (uc) also satisfies
		(k).
	\end{lemma}
	\begin{proof}
		Suppose $C:= R(x,y)\cap R(u,v)\neq \emptyset$ and let $p,q\in C$. By (m),
		$R(p,q)\subseteq C$. If $R(p,q)=C$ there is nothing to show. Otherwise,
		there is $r\in C\setminus R(p,q)$.  By (m), $R(r,q)\subseteq C$ and by
		(uc) there is $p',q'\in R(p,q)\cup R(q,r)$ such that $R(p,q)\cup
		R(q,r)=R(p',q')\subseteq C$.  Furthermore $|R(p',q')|\ge
		|R(p,q)\cup\{r\}|=|R(p,q)|+1$. If $R(p',q')=C$ we are done. Otherwise, we
		replace $p,q$ by $p',q'$ and repeat the argument. This a sequence of
		strictly increasing sets $R(p^k,q^k)\subseteq C$. Since $C$ is finite,
		there is a finite $k$ such that $R(p^k,q^k)=C$, and thus $R$ satisfies
		(k).
	\end{proof}
	
	Union-closed transit functions also satisfy a weak, Helly-like property:
	\begin{lemma}
		\label{lem:triangle}
		Let $R$ be a monotone transit function satisfying (uc) and suppose
		$R(u,v)$, $R(s,t)$, and $R(w,r)$ have pairwisely non-empty
		intersections. Then $R(u,v)\cap R(s,t)\cap R(w,r)\ne \emptyset$.
	\end{lemma}
	\begin{proof}
		Consider three transit sets $A,B,C\in\mathcal{C}_R$ with non-empty
		pairwise intersections. By Lemma~\ref{lem:uc->k}, the intersections
		$E_1:= A\cap B$, $E_2 := B\cap C$ and $E_3:= C\cap A$ are also contained
		in $\mathcal{C}_R$. Since the intersections are non-empty by assumption,
		there are points $x\in E_1$, $y\in E_2$ and $z\in E_3$. Now consider the
		transit sets $R(x,y)$, $R(y,z)$ and $R(z,x)$. By monotonicity, we have
		$R(x,y)\subseteq B$.  Furthermore, $A\cap C\ne\emptyset$ implies that
		$A\cup C\in\mathcal{C}_R$ by (uc) and $x\in A$ and $y\in C$ implies
		$R(x,y)\subseteq A\cup C$ by monotonicity.  Thus $R(x,y)\subseteq
		B\cap(A\cup C)=(B\cap A)\cup(B\cap C)=E_1\cup E_2$.  Similarly, we obtain
		$R(y,z)\subseteq E_2\cup E_3$ and $R(z,x)\subseteq E_3\cup E_1$.  We
		therefore have $R(x,y)\cap R(y,z)\cap R(x,z) \subseteq (E_1\cup E_2) \cap
		(E_2\cup E_3) \cap (E_1\cup E_3)= A\cap B \cap C$.  Thus $A\cap B\cap
		C=\emptyset$ implies $R(x,y)\cap R(y,z)\cap R(x,z)=\emptyset$.
		\newline
		Now consider $D_1:= R(x,y)\cap R(y,z)$, $D_2:= R(y,z)\cap R(z,x)$ and
		$D_3:= R(x,y)\cap R(x,z)$. We have $y\in D_1$, $z\in D_2$ and $x\in D_3$,
		i.e., the three intersections are non-empty. Since (uc) implies (k), we
		have $D_1,D_2,D_3\in\mathcal{C}_R$. From $R(x,y)=D_1\cup D_2$,
		$R(y,z)=D_2\cup D_3$ and $R(z,x)=D_3\cup D_1$ we compute $R(x,y)\cap
		R(x,z)\cap R(y,x)= (D_1\cup D_2) \cap (D_2 \cup D_3) \cap (D_1\cup D_3)$.
		Thus $A\cap B\cap C=\emptyset$ implies that $D_1$, $D_2$, and $D_3$ are
		non-empty and pairwise disjoint.
		\newline
		Since $R(x,y)\cap R(y,z)\ne\emptyset$, (uc) implies that there is $p,q$
		such that $R(p,q)=R(x,y)\cup R(y,z)=D_1\cup D_2\cup D_3$. If $p,q$ are in
		the same same set $D_i$, then $R(p,q)\subseteq D_i\subsetneq R(p,q)$, a
		contradiction.  If $p\in D_i $ and $q\in D_j$ for $i\neq j$ then $R(p,q)$
		is contained in $R(x,y)$ or $R(y,z)$ or $R(z,x)$. This contradicts
		(uc). Therefore $A\cap B\cap C\ne\emptyset$.
	\end{proof}
	
\begin{figure}[t]
		\begin{minipage}{0.40\textwidth}
			\includegraphics[width=\textwidth]{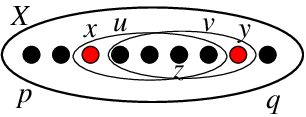}
		\end{minipage} \quad\qquad
		\begin{minipage}{0.50\textwidth}
			\caption{The set system $\mathcal{C}$ on $X$ comprising the singletons,
				$X$, and the three intervals $[x,v]$, $[u,y]$, and $[u,v]$ is clearly
				pyramidal with the linear order $<$ shown from left to right. For its
				canonical transit function $R$ we have $R(x,z)=[x,v]$, $R(z,y)=[u,y]$
				but $R(x,y)=X\ne [x,v]\cup[u,y]=[x,y]\notin\mathcal{C}$. Therefore
				$R$ is pyramidal but satisfies neither (uc) nor (u).}
			\label{fig:counterO1}
		\end{minipage}
	\end{figure}

	The following property appears as a property of cut-vertex transit
	functions of hypergraphs in \cite{Changat:21a} and as property (h'') in
	\cite{Changat:18a} in the context of characterizing hierarchical clustering
	systems:
	\begin{description}\setlength{\itemsep}{0pt}\setlength{\parskip}{0pt}%
		\item[(u)] If $z\in R(u,v)$ then $R(u,v)=R(u,z)\cup R(z,v)$.
	\end{description}
	Figure.~\ref{fig:counterO1} shows, however, that it is not shared by all
	pyramidal clustering systems.
	
	\begin{lemma}\label{lem:uc->u}
		Let $R$ be a monotone transit function. Then (uc) implies (u).
	\end{lemma}
	\begin{proof}
		Consider $z\in R(x,y)$ for some $x,y\in X$. Since
		$R(x,z)\cap R(z,y)\ne\emptyset$, (uc) implies that there are
		$p,q\in R(x,z)\cup R(z,y)$ such that $R(x,z)\cup R(z,y)=R(p,q)$. By
		monotonicity we have $R(x,z)\cup R(z,y)\subseteq R(x,y)$ which implies
		$R(p,q)\subseteq R(x,y)$. Conversely, we have $x,y\in R(p,q)$ and thus
		$R(x,y)\subseteq R(p,q)$, and thus $R(x,y)=R(p,q)$. Thus $z\in R(x,y)$
		implies $R(x,z)\cup R(z,y)=R(x,y)$, i.e., the transit function $R$
		satisfies (u). 
	\end{proof}
	The converse is not true, as the following example \ref{ex:u not uc} shows:
	\begin{example}\label{ex:u not uc}
		Consider the transit function $R$ on $X=\{a,b,c,d\}$ defined by
		$R(a,b)=\{a,b\}, R(a,c)= \{a,c\}, R(b,c)=\{b,c\}, R(a,d) =X$, $R(b,d)=
		R(c,d)=\{b,c,d\}$. Here $R$ is monotone, $R$ satisfies (u) but violates
		(uc) since $\{a,b,c\}=R(a,b)\cup R(a,c)$ is not a transit set.
	\end{example}
	
	The following axiom, which is reminiscent of the triangle inequality,
	appeared in \cite{Changat:21a}:
	\begin{description}\setlength{\itemsep}{0pt}\setlength{\parskip}{0pt}%
		\item[(x)] For any three $x,y,z\in X$ holds $R(x,y)\subseteq R(x,z)\cup
		R(z,y)$.
	\end{description}
	Restricting (x) to those
	triples $x,y,z$ satisfying $z\notin R(x,y)$ relaxes (x) to property (x'),
	which we have introduced earlier.  For $z\in R(x,y)$, the statements of (u)
	and (x) are equivalent, while the statements in (x) and in (x') are the
	same for $z\notin R(x,y)$. Thus we have:
	\begin{lemma}\label{lem:x<->x'+u}
		A monotone transit function $R$ satisfies (x) if and only if it satisfies
		(u) and (x').
	\end{lemma}
	For monotone transit functions, (uc) implies (u) and (w),
	and by Lemma~\ref{lem:w->x'} (w) in turn implies (x'). Therefore
	we have
	\begin{corollary}\label{cor:uc=>x}
		If a monotone transit function $R$ satisfies (uc) then it also
		satisfies (x).
	\end{corollary}

	\begin{definition}
		A set system $(X,\mathcal{C})$ is \emph{union-closed} if it contains all
		singletons and
		\begin{description}\setlength{\itemsep}{0pt}\setlength{\parskip}{0pt}%
			\item[(UC)] $C',C''\in\mathcal{C}$ and $C'\cap C''\ne\emptyset$
			implies $C'\cup C''\in\mathcal{C}$.
		\end{description}
	\end{definition}
	Clearly, a monotone transit function satisfies (uc) if and only if it
	explains a union-closed $\mathcal{T}$-system.
	
\begin{figure}[t]
		\begin{minipage}{0.20\textwidth}
			\includegraphics[width=\textwidth]{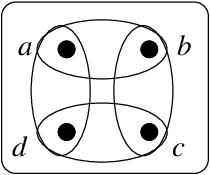}
		\end{minipage} \quad\qquad
		\begin{minipage}{0.70\textwidth}
			\caption{The set system $\mathcal{C}$ on $X$ comprising the singletons,
				$X$, and the four edges $C_1=\{a,b\}$, $C_2=\{b,c\}$, $C_3=\{c,d\}$
				and $C_4=\{d,a\}$ is a weak hierarchy (since any triple of sets that
				intersect pairwise contains either $X$ or a singleton), and satisfies
				axiom (WP). Since the four edges form a 4-cycle $(C_1,C_2,C_3,C_4)$
				in $(X,\mathcal{C})$ there is no linear ordering on $X$ compatible
				with $\mathcal{C}$. Thus $(X,\mathcal{C})$ is weakly (pre-)pyramidal
				but not (pre-)pyramidal.}
			\label{fig:exWP}
		\end{minipage}
	\end{figure}

	Assuming (UC), we have $\mathcal{C}=\langle\mathcal{C}\rangle$. In this
	case, case Nebesk{\'y}'s result becomes ``$\mathcal{C}$ is pre-pyramidal if
	and only if the union closure $\langle\{A,B,C\}\rangle$ is pre-pyramidal
	for all $A, B, C\in \mathcal{C}$". By Proposition \ref{prop:Nebesky2}, this
	happens when $\mathcal{C}$ holds (W') and (WP). We have proved in Lemma
	\ref{lem:1=W} that $\mathcal{C}$ holding (W') is equivalent to
	$\mathcal{C}$ being a weak hierarchy. These results yield a convenient
	characterization of pre-pyramidal set systems:
	\begin{corollary}
		\label{cor:UC+WP=py}
		A weak hierarchy that satisfies (UC) is
		(pre-)pyramidal if and only if it satisfies (WP).
	\end{corollary}
	Example~\ref{ex:pynotuc} shows that there are binary (pre-)pyramidal clustering
	systems that do not satisfy (UC). Hence, requiring (W') and
	(WP) to hold for $\mathcal{C}$ instead of $\langle\mathcal{C}\rangle$
	is only necessary but not a sufficient condition. 
	\begin{definition}
		A clustering system $(X,\mathcal{C})$ is \emph{weakly pyramidal} if it is
		a weak hierarchy and satisfies (WP).
	\end{definition}
	Figure.~\ref{fig:exWP} gives an example of a set system that is
	weakly pyramidal but not pyramidal.
	
	\begin{example}
		Let $R$ be the transit function on $X=\{x,y,z,w\}$ defined by
		$R(x,y)=\{x,y,w\}$, $R(x,z)=\{x,z\}$, $R(x,w)=\{x,w\}$, $R(y,z)=\{y,z\}$,
		$R(y,w)=\{y,w\}$ and $R(z,w)=X$. Then $R$ is monotone and satisfies
		(WP). Here $z\notin R(x,y)$ and $R(x,y)\nsubseteq R(x,z)\cup
		R(z,y)$. Therefore (x') is not satisfied. Also, the sets $R(x,z)$ and
		$R(z,y)$ violate (mm).
	\end{example}
	
	Axiom (WP) can be translated trivially to the language of transit
	functions:
	\begin{description}\setlength{\itemsep}{0pt}\setlength{\parskip}{0pt}%
		\item[(wp)] If $R(u,v)\cap R(x,y)\ne\emptyset$,
		$R(u,v)\cap R(p,q)\ne\emptyset$ and $R(x,y)\cap R(p,q)\ne\emptyset$ then
		$R(p,q)\subseteq R(u,v)\cup R(x,y)$ or
		$R(u,v)\subseteq R(p,q)\cup R(x,y)$ or
		$R(x,y)\subseteq R(p,q)\cup R(u,v)$.
	\end{description}
	The following Example~\ref{ex:W/->WP} shows that (w) does not imply
	(wp) even for monotonous transit functions.
	\begin{example}
		\label{ex:W/->WP}
		For $X=\{a,b,c,d\}$ consider the monotonous transition function defined
		by $R(a,d)=\{a,d\}$, $R(b,d)=\{b,d\}$, $R(c,d)=\{c,d\}$ and $R(x,y)=X$
		for all other pairs $x\ne y$. The the transit sets $\mathcal{C}_R$ form a
		weak hierarchy: Any triple of pairwisely intersecting transit sets
		$\{A,B,C\}$ either contains $X$ or a singleton (in which case
		equation.(\ref{eq:ABC}) is trivially satisfied), or equals $A=R(a,d)$,
		$B=R(b,d)$, $C=R(c,d)$ for which we have
		$A\cap B=A\cap C=B\cap C=A\cap B\cap C=\{d\}$. On the other hand,
		(WP) is clearly violated for $A$, $B$, and $C$.
	\end{example}
	Conversely, (wp) does not imply (w):
	\begin{example}
		For $X=\{a,b,c\}$ consider the ``triangle'' transit function
		$R(a,b)=\{a,b\}$, $R(a,c)=\{a,c\}$, $R(b,c)=\{b,c\}$, which trivially
		satisfies (wp) but violates (w) since $R(a,b)\cap R(a,c)\cap
		R(b,c)=\emptyset$.
	\end{example}
	
	\begin{theorem}
		\label{thm:uc=>wpw}
		Let $R$ be a monotone transit function satisfying (uc). Then
		$R$ satisfies (w) and (wp). 
	\end{theorem}
	\begin{proof}
		Suppose (wp) does not hold. Then there are transit sets $A,B,C$ with
		pairwise non-empty intersections such that $A\nsubseteq B\cup C$,
		$B\nsubseteq A\cup C$ and $C\nsubseteq B\cup A$.  By (uc), there is
		$p,q\in A\cup B\cup C$ such that $A\cup B\cup C=R(p,q)$ for some
		$p,q$. If $p,q$ lie in the same transit set, say $A$, then monotonicity
		implies $R(p,q)\subseteq A\subseteq A\cup B\cup C\subseteq R(p,q)$, i.e.,
		$R(p,q)=A=A\cup B\cup C$ and thus $B\cup C\subseteq A$, contradicting the
		assumption $B\nsubseteq A\cup C$. If $p$ and $q$ belongs to different
		sets, say $A$ and $B$, then $R(p,q)\subseteq A\cup B \subseteq A\cup
		B\cup C=R(p,q)$, and thus $C\subseteq A\cup B$, contradicting the
		assumption $C\nsubseteq B\cup A$. Therefore $A\cup B\cup C\neq R(p,q)$
		for any $p,q$ which contradicts (uc).
		\newline
		Now suppose (w) does not hold. Then there exists $a,b,c\in X$ such that
		$a\notin R(b,c), b\notin R(a,c)$ and $c\notin R(a,b)$. By construction,
		$R(a,b)$, $R(b,c)$, and $R(a,c)$ have non-empty pairwise intersection and
		thus, by Lemma~\ref{lem:uc->k}, the intersections $D_1:= R(a,b)\cap
		R(b,c)$, $D_2:= R(b,c)\cap R(c,a)$, and $D_3=R(a,b)\cap R(a,c)$ are also
		transit sets with $b\in D_1$, $c\in D_2$ and $a\in D_3$. Using (uc) and
		(m) we have $R(a,b)=D_1\cup D_2$, $R(b,c)=D_2\cup D_3$, and
		$R(c,a)=D_3\cup D_1$ and $d\in D_1\cap D_2\cap D_3$. Since $R(a,b)\cap
		R(b,c)\ne\emptyset$ there is $p,q$ such that $R(p,q)=R(a,b)\cup
		R(b,c)=D_1\cup D_2\cup D_3$. If both $p,q$ are in the same set say
		$p,q\in D_1$, then $D_1=R(a,b)=R(p,q)$, contradicting $c\notin
		R(a,b)$. Thus $p$ and $q$ are two distinct sets, say $p\in D_1$ and $q\in
		D_2$. Then $p,q\in R(b,c)$, implying $R(p,q)=R(b,c)$, contradicting
		$a\notin R(b,c)$.
	\end{proof}
	Theorem \ref{thm:uc=>wpw} and Corollary \ref{cor:UC+WP=py} immediatly imply
	\begin{corollary}
		A monotone transit function $R$ that satisfies (uc) is
		(pre)pyramidal.
	\end{corollary}
	
	We note that hierarchies, i.e., clustering systems satisfying
	\begin{description}\setlength{\itemsep}{0pt}\setlength{\parskip}{0pt}%
		\item[(H)] $C',C''\in\mathcal{C}$ and
		$C'\cap C''\ne\emptyset$ implies $C'\subseteq C''$ or $C''\subseteq C'$,
	\end{description}\vspace*{-\topsep}
	are trivially union-closed. On the other hand, $\mathcal{C}=\{ \{1\},
	\{2\}, \{3\}, \{1,2\}, \{2,3\}$, $\{1,2,3\}\}$ is a union-closed binary
	(ucb) clustering system that is clearly not a hierarchy. The class of ucb
	clustering systems defined by axiom (uc) thus are situated properly
	``between'' hierarchies and pyradmids in the sense that among the binary
	clustering systems. Paired hierarchies, also known as 2-3 hierarchical
	clustering, \cite{Chelcea:05,Bertrand:08}, in which a cluster
	$C\in\mathcal{C}$ may properly overlap at most one other cluster, also lie
	between hierarchies and pyramids \cite{Barthelemy:08}. One easily checks
	that ucb clustering systems and paired hierarchies are unrelated: For
	example, $\mathcal{C}_1=\{ \{1\}, \{2\}, \{3\}, \{4\}, \{1,2\}, \{2,3\},
	\{3,4\}, \{1,2,3\}, \{2,3,4\}, \{1,2,3,4\} \}$ clearly is an ucb clustering
	system but not a paired hierarchy. On the other hand $\mathcal{C}_2=\{
	\{1\}, \{2\}, \{3\}, \{4\}, \{1,2\}, \{2,3\}, \{1,2,3,4\} \}$ is a paired
	hierarchy but violates (UC) since $\{1,2,3\}\notin\mathcal{C}_2$.
	
	Two properties related to, but weaker than axiom (u), are:
	\begin{description}\setlength{\itemsep}{0pt}\setlength{\parskip}{0pt}%
		\item[(o)] For all $u,v\in X$ there are $p,q\in R(u,v)$ such that
		$z\in R(u,v)$ implies $R(p,z)\cup R(z,q)=R(u,v)$.
		\item[(o')] For all $u,v\in X$ and $z\in R(u,v)$ there are
		$p,q\in R(u,v)$ such that $R(p,z)\cup R(z,q)=R(u,v)$.
	\end{description}
	Clearly, if (u) holds then (o) is true with $p=u$ and
	$q=v$. Furthermore, (o) trivially implies (o').
	
\begin{figure}[h]
		\begin{minipage}{0.20\textwidth}
			\includegraphics{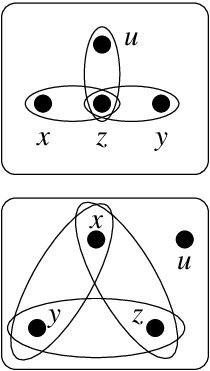} \end{minipage}
		\quad\qquad
		\begin{minipage}{0.70\textwidth}
			\caption{\emph{Top:} The set system $\mathcal{C}$ consists of the
				singletons, the three edges $\{x,z\}$, $\{y,z\}$, $\{u,z\}$, and
				$X$. One easily checks that it is a weak hierarchy. Its canonical
				transit function $R$ satisfies $R(p,q)=X$ if and only if $p\ne q$ and
				$p,q\in\{x,y,u\}$. Obviously $z\in R(p,q)=X$. However, $R(p,z)\cup
				R(z,q)=\{p,q,z\}\ne R(p,q)$; thus $R$ does not satisfy
				(o). \emph{Below:} The set system comprises the singletons, $X$, and
				the three pairs $\{p,q\}$ with $p,q\in\{x,y,z\}$. These three pairs
				intersect pair-wisely, but $\{x,y\}\cap\{x,z\}\cap
				\{y,z\}=\emptyset$, thus $\mathcal{C}$ is not a weak hierarchy.  The
				canonical transit function $R$ satisfies $R(p,q)=\{p,q\}$ for
				$p,q\in\{x,y,z\}$ and $R(p,u)=X$ for $p\in\{x,y,z\}$. For pairs, the
				statement of (o) is trivially true, and for $z\in R(p,u)=X$ we have
				$R(p,u)=R(p,z)\cup R(z,u)$ since $R(z,u)=X$ for $z\ne u$ and
				$R(p,z)=X$ for $z=u$. Thus $R$ satisfies (o) but not (w).}
			\label{fig:WO}
		\end{minipage}
	\end{figure}

	\begin{lemma}\label{lem:py->o}
		Every monotone pyramidal transit function $R$ satisfies (o).
	\end{lemma}
	\begin{proof}
		Let $<$ be a compatible order for the pyramidal cluster system
		$\mathcal{C}_R$ and consider a cluster $R(x,y)\in\mathcal{C}_R$.  Set
		$p:=\min R(u,v)$ and $q:=\max R(u,v)$ (w.r.t.\ $<$). If $z \in R(u,v)$,
		then $p\leq z\leq q$ and thus $z\in R(p,q)$. Therefore $R(u,v)\subseteq
		R(p,q)$. Since $R$ is monotone and $p,q \in R(u,v)$ we have
		$R(p,q)\subseteq R(u,v)$, and hence $R(p,q)=R(u,v)$. Furthermore, $z\in
		R(u,v)$ if and only if $p\le z\le q$, i.e., $R(p,q)=[p,q]$.  Since
		$\mathcal{C}_R$ is pyramidal we have $[p,z]\subseteq R(p,z)$ and
		$[z,q]\subseteq R(z,q)$, and thus $R(p,q)=[p,q]=[p,z]\cup[z,q] \subseteq
		R(p,z)\cup R(z,q)$.  Monotonicity of $R$, on the other hand, implies
		$R(p,z)\cup R(z,q)\subseteq R(p,q)$ for all $z\in R(p,q)=R(x,y)$.
	\end{proof}
	The two examples in Figure.~\ref{fig:WO} show that (o) is independent of
	(w) even for monotone transit functions.
	
	\begin{theorem}
		\label{thm:wp=>o'}
		Let $R$ be a monotone transit function. Then (wp) implies
		(o').
	\end{theorem}
	\begin{proof}
		\par\noindent\textit{Step 1:}
		Consider $R(a,b)$ such that there is $z\in R(a,b)$ with
		$R(a,b)\nsubseteq R(a,z)\cup R(z,b)$. Then there exists $d_1\in R(a,b)$
		such that $d_1\notin R(a,z)$ and $d_1\notin R(z,b)$. Consider the sets
		$R(a,z)$, $R(z,b)$, and $R(z,d_1)$. (wp) implies that:
		Case 1.1: $R(a,z)\subset R(z,b)\cup R(z,d_1)$ or
		Case 1.2: $R(z,b)\subset R(a,z)\cup R(z,d_1)$ or 
		$R(a,d_1)\subseteq R(a,z)\cup R(z,b)$, which is not possible since
		$d_1\notin R(a,z),d_1\notin R(z,b)$.  In Case 1.1,
		$a\in R(z,d_1)\Rightarrow R(a,z) \subset R(z,d_1) \Rightarrow
		R(a,z)\cup R(z,b)\subset  R(d_1,z)\cup R(z,b).$ 
		In Case 1.2, $b\in R(z,d_1)\Rightarrow R(z,b)\subset R(z,d_1)\Rightarrow
		R(a,z)\cup R(z,b)\subset  R(a,z)\cup R(z,d_1).$
		In Case 1.1, we choose $(p_1,q_1)=(d_1,b)$, and in Case 1.2 we choose
		$(a,d_1)$. 
		If $R(a,b)\subseteq R(p_1,z)\cup R(z,q_1)$ we are done. Otherwise we
		proceed as follows:
		\par\noindent\textit{Step 2:}
		If $(p_1,q_1)=(d_1,b)$ then there exists $d_2\in R(a,b)$ such that
		$d_2\notin R(z,d_1)$ and $d_2\notin R(z,b)$. Consider the sets
		$R(z,d_1), R(z,b)$ and $R(z,d_2)$. (wp) implies that:
		Case 2.1: $R(d_1,z)\subset R(z,b)\cup R(z,d_2)$, or Case 2.2:
		$R(z,b)\subset R(z,d_1)\cup R(z,d_2)$, or
		$R(z,d_2)\subseteq R(z,d_1)\cup R(z,b)$, which is not possible since
		$d_2\notin R(z,d_1), d_2\notin R(z,b)$.
		In Case 2.1, $d_1\in R(z,d_2)\Rightarrow R(z,d_1)\subset R(z,d_2)
		\Rightarrow  R(z,d_1) \cup R(z,b) \subset R(z,d_2)\cup R(z,b).$
		In Case 2.2, $R(z,d_1) \cup R(z,b) \subset R(z,d_1)\cup R(z,d_2)$.
		In Case 2.1, we set $(p_2,q_2)=(d_2,b)$,
		in Case 2.2, we set $(p_2,q_2)=(d_1,d_2)$.
		Similarly, if $(p_1,q_1)=(a,d_1)$, then we obtain $(p_2,q_2)=(d_1,d_2)$
		or $(p_2,q_2)=(a,d_2)$. If $R(a,b)\subseteq R(p_2,z)\cup R(z,q_2)$ we are
		done.  Otherwise, repeat the construction. In each step, by construction,
		we choose a $d_i\in R(a,b)$ that is not covered in the previous
		step. Since $R(a,b)$ is finite, no such $d_i$ is available after a finite
		number of steps.  Thus, we get $(p_n,q_n)$ such that
		$R(a,b)\subseteq R(p_n,z)\cup R(z,q_n)$. Thus, there is a pair
		$p,q\in R(a,b)$ such that $R(a,b)= R(p,z)\cup R(z,q)$. Hence (o') is
		satisfied.
	\end{proof}
	
	The converse is not true. The following example \ref{o' not wp} shows that (o') does
	not imply that $\mathcal{C}_R$ satisfies (WP), i.e., that $R$
	satisfies (wp).
	\begin{example}\label{o' not wp}
		Consider the monotone transit function $R$ on $X=\{x,y,z,p,q,r\}$ defined
		by $R(x,y)=\{x,y,p\}=R(x,p)=R(y,p)$, $R(x,z)=\{x,z,r\}=R(x,r)=R(z,r)$,
		$R(y,z)=\{y,z,q\}=R(y,q)=R(z,q)$, and
		$R(x,q)=R(y,r)=R(z,p)=R(p,q)=R(p,r)=R(q,r)=X$. It is not difficult to
		check that $R$ satisfies (o'). However, the three transit sets
		$R(x,y), R(x,z), R(y,z)$ have non-empty pairwise intersections but
		violate property (wp).
	\end{example}
	Furthermore, neither (o') nor (wp) implies (o):
	\begin{example}
		Consider the monotone transit function $R$ on $X=\{a,b,c,d\}$ defined by
		$R(a,b)=\{a,b,c,d\}$, $R(a,c)=\{a,c\}$, $R(b,c)=\{b,c\}$,
		$R(a,d)=\{a,d\}$, $R(b,d)=\{b,d\}$, and $R(c,d)=\{a,c,d\}$. One easily
		checks that $R$ satisfies (o') and (wp), but $R(a,b)$ violates
		(o).
	\end{example}
	The transit function corresponding to the example in Figure.~\ref{fig:exWP}
	is monotone, satisfies (w) and (o'), but violates (o). 
	
	\vspace{1.5cc}
	\section{Concluding remarks}
	
\begin{figure}[h]
		\begin{center}
			\includegraphics[width=0.7\textwidth]{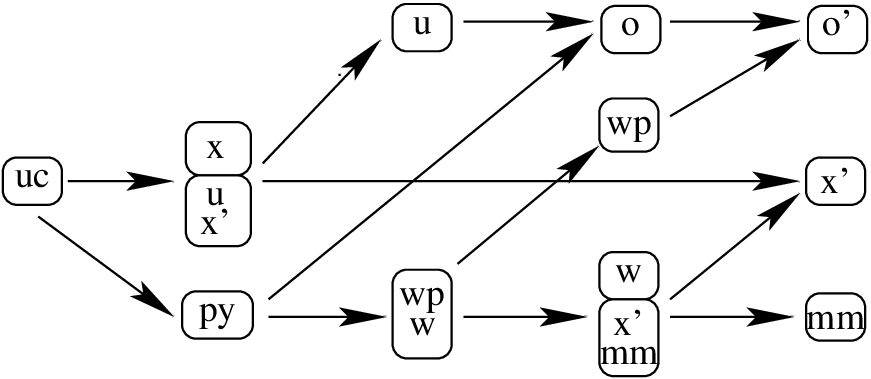}
		\end{center}
		\caption{Summary of implications between properties of monotonous transit
			functions proved in this contribution. (py) indicated a monotonous
			transit function whose transit sets are pyramidal.}
		\label{fig:sum}
	\end{figure}

	This contribution extends previous investigations into the connection
	between transit functions and binary clustering systems. First, we
	revisited weak hierarchies and proved additional characterizations for the
	clustering system and their identifying transit functions.Then, we focused
	on several properties that are related to pyramidal set systems or,
	equivalently, interval hypergraphs. It remains an open question whether
	monotonous transit functions whose transit sets form a pyramidal clustering
	system can be characterized by a finite set of first order axioms.  In
	particular, we identified union-closed set systems and equivalently
	monotonous transit functions satisfying axiom (uc) as an interesting
	restriction of pyramids that is still more general than hierarchies. Our
	results also showed some necessary conditions for pyramidality that define
	interesting subclasses of weak hierarchies. Figure.~\ref{fig:sum}
	summarizes the implications established in this contribution.
	
	Axiom (u), in particular, has appeared already in earlier work on
	characterizations of hierarchical clustering systems \cite{Changat:18a}.
	It remains an interesting open question whether weak hierarchies satisfying
	(u) are pyramidal (py) or at least weakly pyramidal
	(w)$\wedge$(wp). Properties of less-well-studied clustering systems such as
	2-hierarchies \cite{Jardine:71,Barthelemy:04} are also worth studying in
	this context. \\

	\noindent\textbf{Acknowledgements}\\
	\textit{This research work was performed in part while MC was visiting the
		Max Plank Institute for Mathematics in the Sciences (MPI-MIS), Leipzig
		and Leipzig University's Interdisciplinary Center for Bioinformatics
		(IZBI). MC acknowledges the financial support of the MPI-MIS, the
		hospitality of the IZBI, and the Commission for Developing Countries of
		the International Mathematical Union (CDC-IMU) for providing the
		individual travel fellowship supporting the research visit to
		Leipzig. This work was supported in part by SERB-DST, Ministry of Science
		and Technology, Govt.\ of India, under the MATRICS scheme for the
		research grant titled ``Axiomatics of Betweenness in Discrete
		Structures'' (File: MTR/2017/000238).}

	\bigskip
	
	\noindent\textbf{Manoj Changat}\\
	University of Kerala\\
	Department of Futures Studies\\
	Trivandrum 695 581, India\\
	E-mail: {\it mchangat@keralauniversity.ac.in}\\
	\url{https://orcid.org/0000-0001-7257-6031}
	
	\vspace{0.1cc}

	\noindent \textbf{Ameera Vaheeda Shanavas}\\
	University of Kerala\\
	Department of Futures Studies\\
	Trivandrum 695 581, India\\
	E-mail: {\it ameerasv@gmail.com}\\
	\url{https://orcid.org/0000-0003-0038-5492}
	
	\vspace{0.1cc}
	
	\noindent \textbf{Peter F.\ Stadler}\\
	Universit{\"a}t Leipzig\\
	Bioinformatics Group, \\Department of Computer Science and
	Interdisciplinary Center for Bioinformatics,\\ 
	H{\"a}rtelstrasse 16-18, D-04107 Leipzig, Germany\\
	E-mail: {\it studla@bioinf.uni-leipzig.de}\\
	\url{https://orcid.org/0000-0002-5016-5191}\\
	\par\vspace*{-0.6\baselineskip}\noindent
	Max Planck Institute for Mathematics in the Sciences\\
	Leipzig, Germany\\
	\par\vspace*{-0.6\baselineskip}\noindent
	University of Vienna\\
	Institute for Theoretical Chemistry\\
	Vienna, Austria\\
	\par\vspace*{-0.6\baselineskip}\noindent
	Universidad Nacional de Colombia\\
	Facultad de Ciencias\\
	Bogot{\'a}, Colombia\\
	\par\vspace*{-0.6\baselineskip}\noindent
	Santa Fe Institute\\
	Santa Fe, NM
	
\end{document}